\documentclass[12pt]{iopart}
\usepackage{iopams}  
\usepackage{graphicx}

\begin{document}

\title[Geometry of adiabatic Hamiltonians]{Geometry of adiabatic Hamiltonians for two-level quantum systems}

\author{J M S Lehto and K-A Suominen}

\address{Department of Physics and Astronomy, Turku Centre for Quantum Physics and Laboratory of Quantum Optics, University of Turku, FI-20014 Turku, Finland}
\ead{jaakko.lehto@utu.fi}
\vspace{10pt}
\begin{indented}
\item[]\today
\end{indented}

\begin{abstract}
We present the formulation of the problem of the coherent dynamics of quantum mechanical two-level systems in the adiabatic region in terms of the differential geometry of plane curves. We show that there is a natural plane curve corresponding to the Hamiltonian of the system for which the geometrical quantities have a simple physical interpretation. In particular, the curvature of the curve has the role of the nonadiabatic coupling.
\end{abstract}

% Uncomment for PACS numbers
%\pacs{00.00, 20.00, 42.10}
%
% Uncomment for keywords
%\vspace{2pc}
%\noindent{\it Keywords}: XXXXXX, YYYYYYYY, ZZZZZZZZZ
%
% Uncomment for Submitted to journal title message
%\submitto{\JPA}
%
% Uncomment if a separate title page is required
%\maketitle
% 
% For two-column output uncomment the next line and choose [10pt] rather than [12pt] in the \documentclass declaration
%\ioptwocol
%

\section{Introduction}

Two-level quantum systems (TLS), being the simplest of nonsimple quantum systems \cite{Berry263}, can be used to study a wealth of phenomena in quantum physics. Basically, it describes systems with two discrete states, for example the spin degrees of freedom of spin-$\case{1}{2}$ particle, but is often used to describe systems only effectively, as in quantum optics where it serves as a model for resonant excitation of an atom \cite{Shore2011}. More recently, with the advent of the field of quantum information, understanding the dynamics of TLS that act as the quantum information carriers and are therefore dubbed qubits, has become even more important in order to both understand the fundamental physics and gain control over the information processing tasks  \cite{Stenholm2005}. 

Considering modern physics more generally, the appreciation of the fact that the physical phenomena should not depend on the particular coordinate description employed by a physicist has become one of the central principles and has lead to increasingly refined and abstract geometrical tools and language with which to express the physical laws \cite{chruscinski}. In contrast, in this paper we shall discuss the dynamics of a driven TLS in a more elementary differential geometrical setting, but nevertheless try to exploit fully the geometrical character of the resulting theory.

Coherent driving of the system with classical external fields and parameters leads to time-dependencies in the matrix elements of the system Hamiltonian. We can take the Hamiltonian to be real symmetric and traceless without any loss in generality \cite{Berry201}. This means that a particular model, i.e. external driving in question, is described in general by two different functions of time. These are usually called as the detuning and the Rabi frequency, stemming from the quantum optical setting, but they can have any physical origin, such as the two independent spatial components of the external magnetic field in the case of a spin-$\case{1}{2}$ particle. The models where the evolution of the state vector is exactly solvable are rare, despite the long history of the problem \cite{Zener1932, Landau1932, Stuckelberg1932, Majorana1932}. From a mathematical point of view, a TLS Hamiltonian, or the detuning and Rabi frequency defining it, can be thought of as defining a plane curve with time variable as its parameter. We show that there is a natural way of defining the plane curve from these functions in a way that the relations between the resulting geometrical quantities and the basic physical variables conventionally associated with the TLSs exhibit certain simplicity. Despite the elementary character of the results, these are not usually explicitly discussed in treatments of quantum dynamics of TLS. There are many interesting papers \cite{Rojo2010, Berry242, Berry206} in a similar vein that relate the Hamiltonian of the TLS and the geometry of plane or space curves but the way the relation is done  differs from our approach. 

The paper is organised as follows. In the section \ref{sec:formalism} we first give the mathematical formalism and definitions used to describe coherently driven TLS in \ref{subsec:TLS} and connect the theory with the differential geometry of plane curves in \ref{subsec:planecurves}. We give some examples of the usefulness of this connection in subsection \ref{subsec:examples}. The discussion in \ref{sec:discussion} will conclude this presentation. The basic mathematical results of plane curves are given as an Appendix. 

\section{Mathematical Formalism}
\label{sec:formalism}

\subsection{Two-level Quantum Systems}
\label{subsec:TLS}

Generally, the coherent time evolution of a quantum system is given by the Schr{\"o}dinger equation, in units where $\hbar = 1$,

\begin{equation}
\rmi \partial_{t}\bi{\psi}(t) = H(t)\bi{\psi}(t),
\label{eqn:scheq}
\end{equation}
where $H(t)$ is the Hamiltonian operator with time-dependent components, acting on the Hilbert space of the system. For TLS, the space consists of two-dimensional complex state vectors $ \bi{\psi} (t) = \left( c_{+}(t), c_{-}(t) \right)^{T} $, where $c_{+}(t)$ and $c_{-}(t)$ are the complex probability amplitudes of the states formed by the natural basis of the vector space. Physically these basis states, denoted by $\varphi_{\pm}$, usually refer to the two system state vectors in the absence of interaction and the basis is usually called bare or diabatic. In this basis the Hamiltonian is denoted by $H_{d}$ and takes the form
\begin{eqnarray}
H_{d}(t) &= \bi{H}(t)\cdot\bi{\sigma} \\ \nonumber
		 &= \left( \begin{array}{lr}
			\Delta (t) & \Omega(t) \\
			\Omega (t) & -\Delta (t)
					\end{array}\right), 
\label{eqn:hbare}		 		
\end{eqnarray}
where the first line contains the shorthand form involving the field vector $\bi{H}(t) = \left(\Omega(t), 0, \Delta(t) \right)^{T} $ and the Pauli matrices. The function $\Delta(t)$ is called detuning and $\Omega(t)$ is called the Rabi frequency. This is the most general form needed for TLS, as any Hermitian two-by-two matrix can be transformed into this form by suitably redefining the phases of the basis vectors \cite{Berry201}. Instead of the field vector given in the cartesian coordinates, it is also useful to consider the Hamiltonian in the spherical coordinates, in which case
\begin{eqnarray}
H_{d}(t) =  \rho(t)\left( \begin{array}{lr}
			\cos(\theta(t)) & \sin(\theta(t)) \\
			\sin(\theta(t)) & -\cos(\theta(t))
					\end{array}\right), 
\label{eqn:hspherical}					
\end{eqnarray}
where $\rho(t) = \sqrt{\Delta^{2}(t) + \Omega^{2}(t)}$ is the length of the field vector and $\theta(t)$ is the angle it makes with the {\it z}-axis, so that $\Omega/\Delta = \tan (\theta)$. The eigenvalues of the Hamiltonian, the quasi-energies, are given by 
\begin{equation}
E_{\pm}(t) = \pm \rho(t).
\label{eqn:eigenvalues}
\end{equation}

The corresponding eigenstates of the Hamiltonian (\ref{eqn:hbare}) and (\ref{eqn:hspherical}), which can be chosen to be real, read

\begin{equation}
\chi_{+}(t) = \pm \left( \begin{array}{c} \cos ( \frac{\theta(t)}{2} ) \\
									  \sin	( \frac{\theta(t)}{2}) \end{array} \right) , \qquad \chi_{-}(t) = \pm \left( \begin{array}{c} -\sin ( \frac{\theta(t)}{2}) \\
										   \cos	( \frac{\theta(t)}{2} ) \end{array} \right).
\end{equation}
The adiabatic basis, which is formed by these time-dependent eigenvectors (choosing sign to be the same for both eigenvectors), is obtained by $\chi_{\pm} = \bi{U}\varphi_{\pm}$ where the time-dependent unitary transformation is given by
\begin{equation}
\bi{U}(t) =  \left( \begin{array}{lr}
			\cos(\frac{\theta(t)}{2}) & -\sin(\frac{\theta(t)}{2}) \\
			\sin(\frac{\theta(t)}{2}) & \cos(\frac{\theta(t)}{2})
					\end{array}\right).
\label{eqn:basischange}					
\end{equation}

The Schr{\"o}dinger equation in the adiabatic basis reads
\begin{eqnarray}
\rmi \frac{\rmd}{\rmd t}\left( \begin{array}{c} a_{+}(t) \\ a_{-}(t) \end{array}\right) = \left(\begin{array}{lr} \rho(t) & \rmi \gamma (t) \\
																						-\rmi \gamma (t) & -\rho(t)	\end{array}\right) 
																						\left(\begin{array}{c} a_{+}(t) \\ a_{-}(t)  \end{array}\right),
\label{eqn:matrixadiabscheq}
\end{eqnarray}
where $\bi{\psi} = a_{+}\bi{\chi}_{+} + a_{-}\bi{\chi}_{-}$ and the fact that the transformation is time-dependent induces a gauge term in the adiabatic Hamiltonian that couples the adiabatic basis states. This term is called adiabatic coupling $\gamma (t)$ and is given by 
\begin{eqnarray}
\gamma(t) &\equiv -\langle \chi_{+} \vert \dot{\chi}_{-}(\tau) \rangle \nonumber \\ 
&= \frac{\Delta(t)\dot{\Omega}(t) - \dot{\Delta}(t)\Omega(t)}{2\left(\Delta^{2}(t) + \Omega^{2}(t) \right)} \nonumber \\
&= \frac{\dot{\theta}(t)}{2}, 
\label{eqn:acoupl}
\end{eqnarray}
where the overhead dot stands for time derivation. When the Hamiltonian changes slowly, it is well known that the transitions between the adiabatic states vanish and the evolution is called adiabatic. We take $ a_{-}\left( t_{i} \right) = 1$ as the initial condition and define the final transition probability as $P = \vert a_{+}\left(t_{f}\right) \vert^{2} $ where $t_{i}$ and $ t_{f}$ refer to initial and final times, respectively. One can often take $t_{i} = -\infty$ and $t_{f} = +\infty$ but we consider also finite initial and final times. The condition for adiabatic evolution is given by \cite{Shore2011}
\begin{equation}
\vert \Delta(t)\dot{\Omega}(t) - \dot{\Delta}(t)\Omega(t) \vert \ll \left[ \Delta^{2}(t) + \Omega^{2}(t) \right]^{3/2}.
\label{eqn:adiabaticitycondition}
\end{equation}

We consider here the generic case where the functions $\Delta(t)$ and $\Omega(t)$ have different zeros, so there is only avoided crossings in adiabatic energies. However, the diabatic energy levels, given by $\pm \Delta(t)$, can cross, a fact that is exploited in the rapid adiabatic passage techniques \cite{Vitanov2001}. A physically relevant assumption is that the Rabi frequency that couples the diabatic states vanishes when $t  \rightarrow \pm \infty$. Then, or more generally, when the condition
\begin{equation}
\vert \Omega(t) \vert \ll \vert \Delta (t)  \vert, \quad  t \rightarrow t_{i, f}, 
\label{eqn:asymptoticbasis}
\end{equation}
is satisfied, the adiabatic and diabatic basis coincide at the initial and final times. However, one should note that depending on the character of the detuning function, the diabatic and adiabatic basis states may swap labels in relation to each other. That is, it can happen that $\chi_{-}(t_{f}) = \varphi_{+}$ although $\chi_{-}(t_{i}) = \varphi_{-}$, and in this case the transition from the state $\varphi_{-}$ to state $\varphi_{+}$ can be obtained with an adiabatic time evolution. In spherical coordinate description, this is obtained when
\begin{equation}
\theta(t_{f}) - \theta(t_{i}) \equiv 2\int_{t_{i}}^{t_{f}} \gamma(t) \rmd t = 2\pi\left( n + \frac{1}{2}\right),
\label{eqn:deltatheta}
\end{equation}
where $n$ is an integer and the bases coincide in the initial time when we choose $\theta(t_{i}) = 0$. Other interesting cases then are $\theta(t_{f}) = 2\pi n $ and $\theta(t_{f}) =  \pi\left( n + \frac{1}{2}\right)$ which correspond to complete population return and formation of equal superposition, respectively, in the adiabatic evolution.

It should be also noted that the assumption of the non-crossing adiabatic levels allows us to define the change of the time variable, 
\begin{equation}
s(t) = \int_{0}^{t}\rho(x)dx,
\label{eqn:naturaltimescale}
\end{equation}
which means that the time is measured in terms of the dynamical phase that is accumulated. This transformation of the time variable proves to be very important in what follows, and it has the effect of normalizing the field vector in (\ref{eqn:hbare}) and the eigenenergies (\ref{eqn:eigenvalues}), which become $E_{\pm}(s) = \pm 1$. Therefore, we could  choose the detuning and the Rabi frequency as 
\begin{eqnarray}
\Delta(s) &= \cos(\theta(s)) \\
\Omega(s) &= \sin(\theta(s)),
\label{eqn:deltaomega}
\end{eqnarray}
which shows that there is only one function needed to describe any model of coherent TLS dynamics, namely $\theta(s)$.

\subsection{Connection to Plane Curves}
\label{subsec:planecurves}

A curve considered as a connected set of points in plane is a very intuitive concept and one can easily picture it in ones mind. Therefore, it is worthwhile to connect the driven TLS problem of the previous subsection with some suitably defined plane curves. To this end, it is convenient to consider explicitly a parametrized plane curve $\bi{\alpha}(t) = \left( x(t), y(t)\right)$. The basic results from the theory of differential geometry of plane curves that are  needed in the following are given, along with the further references, in the Appendix. We choose to associate with each diabatic Hamiltonian (\ref{eqn:hbare}) a plane curve whose components are given by
\begin{equation}
x(t) = \int_{0}^{t} \Delta (u) du, \quad y(t) = \int_{0}^{t} \Omega (u) du,
\label{eqn:xycurve}
\end{equation}
the parameter $t $ being the physical time and usually we take $t \in \left( -\infty, +\infty \right)$. %The interval $I$ is usually taken to be $\left( -\infty, +\infty \right)$ but sometimes also $ \left( 0, +\infty \right)$ or $ $ 
Note that the so-called Hamiltonian curves defined by Berry in \cite{Berry206} and used to study driven quantum systems are the velocity, or the (unnormalized) tangent, vectors to the curves defined by (\ref{eqn:xycurve}). With our definition, the curvature of the curve is simply related to the functions in the Hamiltonian and given by
\begin{equation}
\kappa[\bi{\alpha}](t) = \frac{2\gamma (t)}{\rho (t)},
\label{eqn:curvature2}
\end{equation}
as is obvious when we look at the formulas (\ref{eqn:acoupl}) and (\ref{eqn:curvature1}).  The adiabatic coupling is directly proportional to curvature and  the adiabatic condition (\ref{eqn:adiabaticitycondition}) can be simply translated to
\begin{equation}
\kappa[\bi{\alpha}](s) \ll 1.
\label{eqn:curvaturecondition}
\end{equation}

Furthermore, the speed of the curve is now just given by $\rho(t)$ and the existence of only avoided crossings means that the curve (\ref{eqn:xycurve}) is regular and its unit-speed parametrization is given by the change of the time variable (\ref{eqn:naturaltimescale}). In this parametrization, the curvature and its relation to the adiabatic coupling simplifies further as the denominators in (\ref{eqn:curvature2}) and (\ref{eqn:curvature1}) become unity. Then from (\ref{eqn:deltaomega}) it follows that curvature is given by
\begin{equation}
\kappa[\bi{\alpha}](t) = \frac{\rmd \theta (s)}{\rmd s},
\label{eqn:curvature3}
\end{equation}
which shows that the angle in (\ref{eqn:deltaomega}) is the same as the so-called turning angle of the curve (\ref{eqn:xycurve}), up to a constant (see figure \ref{fig:ex1}).

\subsection{Applications and examples}
\label{subsec:examples}

\begin{figure}[hb]
\begin{center}
\includegraphics[scale=0.6]{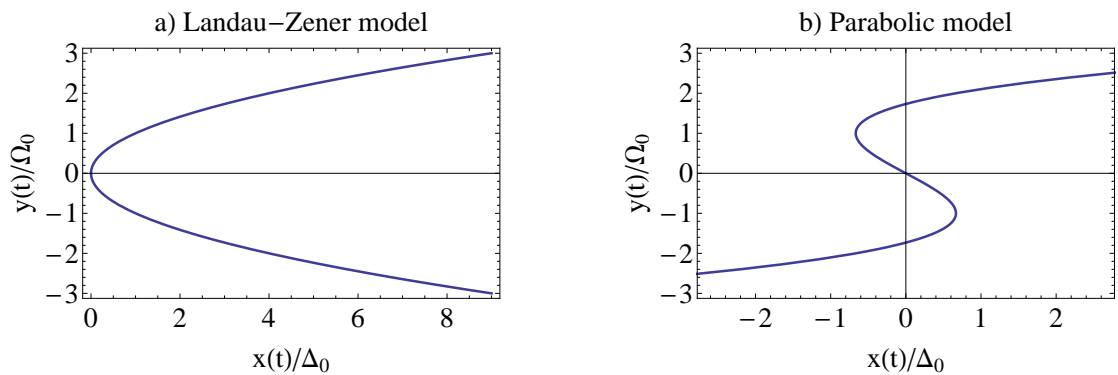} 
\caption{Examples of plane curves defined by TLS model Hamiltonians when $t \in (-3, 3)$: a) Landau-Zener model, $\Delta(t) =\Delta_{0} t $, $\Omega(t) = \Omega_{0}$, b) The double-crossing parabolic model, $\Delta(t) =\Delta_{0} \left(t^{2} - 1 \right) $, $\Omega(t) = \Omega_{0}$ where $\Delta_{0}$ and $\Omega_{0}$ are constants. }
\label{fig:ex1}
\end{center}
\end{figure}

The curves that are formed from the known soluble models are typically very simple and are not bounded in any finite region of the plane. This is due to fact that we usually have $t_{i} = -t_{f} = -\infty$ while the adiabatic coupling, and therefore the corresponding curvature, differ appreciably from zero only near avoided crossings (which are usually chosen to happen near $t=0$) and so the curves have well-defined lines as their asymptotes. Figure \ref{fig:ex1} depicts the situation for Landau-Zener \cite{Zener1932, Landau1932, Stuckelberg1932, Majorana1932} and parabolic \cite{SuominenParabolic, LehtoSuominen2012, Lehto2013} models. It also shows whether or not the model swaps the labels between the diabatic and adiabatic states, as discussed in connection with (\ref{eqn:deltatheta}).

\begin{figure}[hb]
\begin{center}
\includegraphics[scale=0.6]{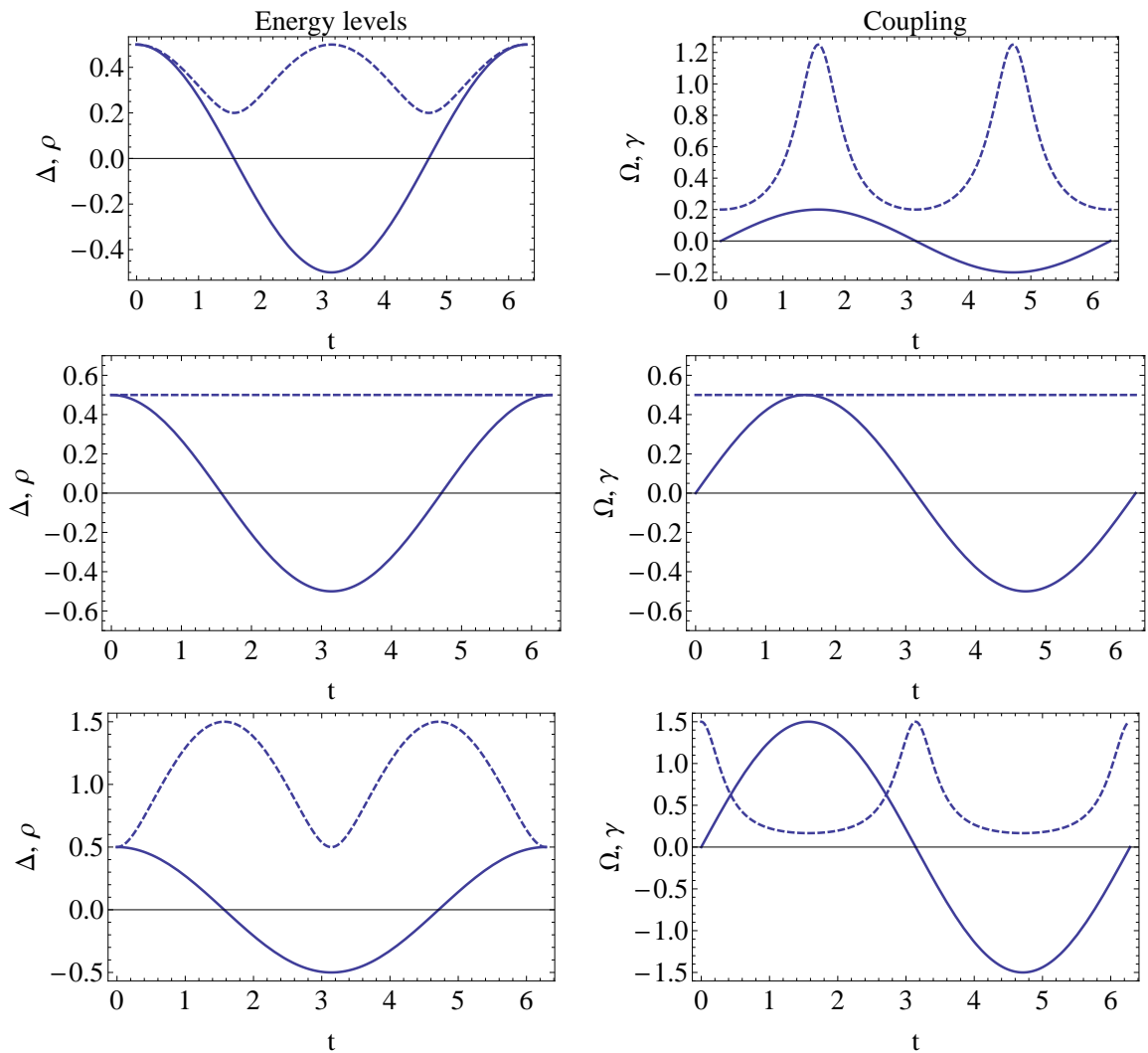} 
\caption{Examples of the diabatic (full line) and the adiabatic (dashed) energy level structure of the Hamiltonian defined by the ellipse (\ref{eqn:ex1ellipse}) are plotted on the left panel while the corresponding couplings are depicted on the right panel. In the uppermost case we have  $\Delta_{0} \geq \Omega_{0}$ and $\Omega_{0} = 0.2$, in middle panel we have the case of a circle with radius $\Omega_{0} = 0.5$ and in the lowermost picture we have $\Omega_{0} \geq \Delta_{0}$, $\Omega_{0} = 1.5$. In all of the plots $\Delta_{0} = 0.5$.}
\label{fig:ellipsefunctions}
\end{center}
\end{figure}

However, we can take the opposite point of view and consider some well-known plane curves with interesting geometric properties from the beginning and then  study what kind of Hamiltonians they give rise to. For example, the ellipse is given by parametric equations
\begin{eqnarray}
x_{1}(t) &= \frac{\Delta_{0}}{\omega}\sin(\omega t) \\
y_{1}(t) &= \frac{\Omega_{0}}{\omega}\left(1 - \cos(\omega t)\right),
\label{eqn:ex1ellipse}
\end{eqnarray}
where $\Delta_{0}$, $\Omega_{0}$ and $\omega$ are positive constants. The way the ellipse is defined here places it completely on the positive half of the y-plane and the respective magnitudes of the constants $\Delta_{0}$ and $\Omega_{0}$ determine the minor and major axes, i.e., whether the ellipse is squeezed in $x_{1}$- or $y_{1}$-direction. %in which direction the ellipse is squeezed. 
The corresponding time-dependent Hamiltonian can be described by the field vector in the $\left( \bi{e}_{1}, \bi{e}_{2}, \bi{e}_{3}\right)$-coordinate system
\begin{equation}
\bi{H}_{1}(t) = \left(\Omega_{0}\sin(\omega t), 0, \Delta_{0}\cos(\omega t) \right)^{T},
\label{eqn:fieldvectorEllipse}
\end{equation}
which also gives an ellipse rotating in $\bi{e}_{1}$-$\bi{e}_{3}$ plane. Although the angular frequency $\omega$ can be a physical parameter, it does not affect the geometrical properties, so we set it equal to unity. In particular, one should note that the value of $\omega$ does not affect the adiabaticity (see equation (\ref{eqn:curvature1})). We furthermore restrict our time to the one-period interval $t \in \left[0, 2\pi \right]$. This diabatic Hamiltonian is then an example of zero-pulse model with two level crossings at $t = \frac{\pi}{2}$ and $t = \frac{3\pi}{2}$ with any parameter values of $\Delta_{0}$ and $\Omega_{0}$. The adiabatic and diabatic bases coincide at initial and final times. This can, of course, be obtained directly from (\ref{eqn:fieldvectorEllipse}) but a nice way is also to imagine the ellipse (\ref{eqn:ex1ellipse}) and to note the connection between the function $\theta(s)$, defining the Hamiltonian in (\ref{eqn:deltaomega}), and the turning angle, which makes it immediately obvious that $\theta(0) = \theta(2\pi) = 0$.  
The curvature of the original curve (\ref{eqn:ex1ellipse}) is given by
\begin{equation}
\kappa_{1}(t) = \frac{\Delta_{0}\Omega_{0}}{\left(\Delta_{0}^{2}\cos^{2}(t) + \Omega_{0}^{2}\sin^{2}( t)  \right)^{3/2}},
\label{eqn:kappa1}
\end{equation}
and the speed of the curve is given by 
\begin{equation}
\rho_{1}(t) = \sqrt{\Delta_{0}^{2}\cos^{2}(t) + \Omega_{0}^{2}\sin^{2}(t)}.
\end{equation}
The latter, of course, gives the eigenenergies of the Hamiltonian and also the adiabatic coupling is directly read from these expressions via (\ref{eqn:curvature2}). The corresponding quantities are plotted in figure \ref{fig:ellipsefunctions}.

The basic connection made between the plane curves and Hamiltonians in \ref{subsec:planecurves}, that was based on a simple observation about the similarities between the mathematical expressions of different quantities, may seem rather trivial. The point, however, is that now one can make use of the many strong  differential geometric results for plane curves. One of the most celebrated results, of global character, is the four-vertex theorem (FVT) \cite{GrayBook} which says that the curvature of a simple closed curve (a circle exluded) has to have in total at least four points of local maxima and minima. Because of the association of the nonadiabatic coupling as curvature and the fact that multiple peaks in the coupling can lead to interference effects because of the distinct times when the transition happens, it is interesting to study the implications of FVT, namely, that for certain type of models the distinct peaks in the adiabatic coupling can not be made to go away.

As a simple plane curve, the ellipse obeys the FVT and it has exactly four vertices. at $t = 0, \frac{\pi}{2}, \pi, \frac{3\pi}{2}$. Which ones correspond to maximas and which minimas, depends on the parameters $\Delta_{0}$ and $\Omega_{0}$, but in any case $\kappa (0) = \kappa (\pi)$ and $\kappa (\frac{\pi}{2}) = \kappa (\frac{3\pi}{2})$ and these values are 
\begin{eqnarray}
\kappa_{1} (0) &= \frac{\Omega_{0}}{\Delta_{0}^{2}}, \\
\kappa_{1} (\frac{\pi}{2}) &= \frac{\Delta_{0}}{\Omega_{0}^{2}}.
\label{eqn:vertices}
\end{eqnarray}
The adiabatic limit of this model is not immediately obvious. If $\Omega_{0} \geq \Delta_{0}$, the maxima of curvature are at points $t = 0, \pi $ and minima at $t = \frac{\pi}{2}, \frac{3\pi}{2}$. These maximum peaks become higher and more narrow when the ratio $\Omega_{0}/\Delta_{0}$ gets bigger. On the other hand, keeping $\Omega_{0}$ fixed, these maxima of curvature become smaller when we increase the value of $\Delta_{0}$. However, this has the effect of increasing the value of the curvature at the minimum points. Indeed, when $\Delta_{0} = \Omega_{0}$, the curvature is a constant function with non-zero value $\Delta_{0}^{-1}$ as the curve corresponds to a circle of radius $\Delta_{0}$. The adiabatic limit for the TLS defined by this circle is obtained as its radius tends to infinity. Also for a general ellipse we see that the adiabatic condition (\ref{eqn:curvaturecondition}) is reached, for example, if we increase both $\Omega_{0}$ and $\Delta_{0}$ but keep the ratio $\Omega_{0}/\Delta_{0}$ constant. The case when we further increase the value of $\Delta_{0}$,  so that $\Delta_{0} \geq \Omega_{0}$, is basically similar to the case $\Omega_{0} \geq \Delta_{0}$, but now the maxima and minima of the curvature have swapped places. This is particularly interesting when we compare the evolution in diabatic and adiabatic bases. 

In diabatic basis, the basic structure of the energy levels and corresponding couplings remains the same in all parameter regions. It has two level-crossings at fixed times when the coupling is also maximal, while the coupling is zero in the initial and final times, see figure \ref{fig:ellipsefunctions}. In adiabatic basis, we have to consider different parameter regions. Although the bases coincide at initial and final times in all parameter regions, the structure of the adiabatic levels and couplings depend on the respective magnitudes of $\Delta_{0}$ and $\Omega_{0}$. When $\Delta_{0} \geq \Omega_{0}$ the avoided crossings of the levels happen at the same instant as the crossings in the diabatic basis and also the adiabatic coupling is maximal there. So in both bases the transitions are located near these two points. When $\Omega_{0} \geq \Delta_{0}$, the maxima of the curvature swap with the minima as discussed above, and there is only one avoided crossing and a peak in the adiabatic coupling between $t_{i}$ and $t_{f}$ at $t = \pi$. However, the adiabatic coupling obtains its maximum also at the initial and final times.

\begin{figure}[hb]
\begin{center}
\includegraphics[scale=0.6]{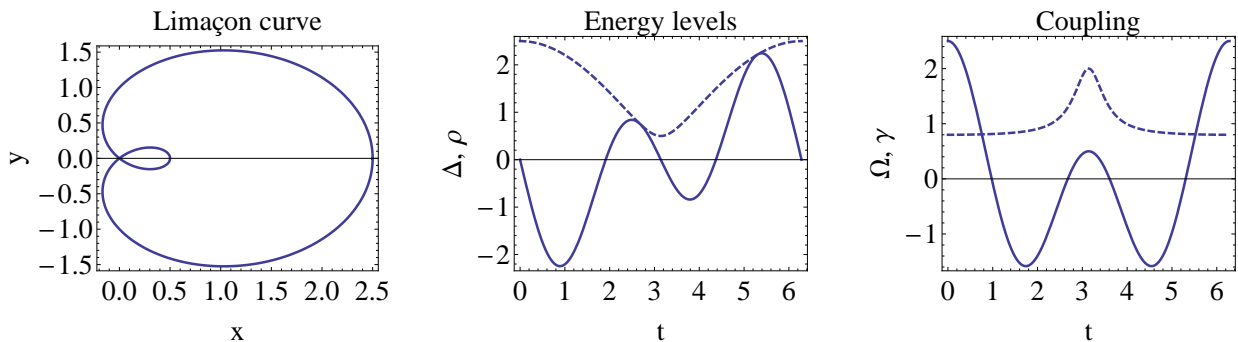} 
\caption{Left: Lima\c{c}on curve of equation (\ref{eqn:limacon}) with $a = 1.5$, $b = 1$. Middle: Energy levels corresponding to Hamiltonian (\ref{eqn:Hamiltonian2}). Diabatic levels are depicted in full line while the adiabatic levels are dashed. Right: The diabatic (full line) and the adiabatic (dashed) coupling of (\ref{eqn:Hamiltonian2}).   }
\label{fig:limacon}
\end{center}
\end{figure}

As another example we can consider the lima\c{c}on curve which is not simple so that FVT is not valid in this case. It is given in the parametric form by
\begin{eqnarray}
x_{2}(t) &= \left(a \cos(\omega t) + b \right)\cos(\omega t) \\ 
y_{2}(t) &= \left(a \cos(\omega t) + b \right)\sin(\omega t),
\label{eqn:limacon}
\end{eqnarray}
where $t \in \left[0, 2\pi \right]$ and we again have three parameters $a$, $b$ and $\omega$, all appearing now in both components. The latter parameter is set to unity for the same reason as in the previous example. Of course there are special curves defined by (\ref{eqn:limacon})  which correspond to special parameter values. For example, when either $a=0$ or $b=0$ we have a circle and the considerations of the adiabatic limit in the previous example apply. On the other hand, if $a = b$, the equations (\ref{eqn:limacon}) define a cardioid, which has a singularity in the curvature function at $t = \pi$, and so we cannot expect to obtain an adiabatic evolution over the whole curve. The curve (\ref{eqn:limacon}) gives rise to a Hamiltonian with the detuning function and Rabi frequency as
\begin{eqnarray}
\Delta_{2}(t) &= -b\sin(t) - a\sin(2t) \\
\Omega_{2}(t) &= b\cos(t) + a\cos(2t).
\label{eqn:Hamiltonian2}
\end{eqnarray}
Curvature and speed in the case are given by
\begin{eqnarray}
\kappa_{2}(t) &= \frac{2a^{2} + b^{2} + 3 a b \cos(t)}{(a^{2} + b^{2} + 2 a b \cos(t))^{3/2}}, \\
\rho_{2}(t) &=\sqrt{a^{2} + b^{2} + 2 a b \cos(t)}.
\end{eqnarray}
We consider the case where $a > b > 0$, so that the curve is not simple. It is still regular and has an inner loop, see figure \ref{fig:limacon}. The FVT does not apply now and curvature has only one maximum, at $t = \pi$ and 
\begin{equation}
\kappa_{2}(\pi) = \frac{1}{a - b}\left( 1 + \frac{a}{a - b}\right),
\end{equation}
and the adiabatic limit is obtained when $\vert a - b \vert$ tends to infinity. 

We have chosen for convenience the initial time as $t_{i} = 0$ but of course any one-period time interval, i.e. $T = 2\pi$, would suffice to trace the lima\c{c}on curve. With the current convention, the diabatic levels start at degeneracy and with maximal diabatic coupling. A similar  discrepancy between the description in different bases that was met in the first example is also present here (see figure \ref{fig:limacon}). In adiabatic basis, there is only one avoided crossing, at $t = \pi$, and the corresponding peak in the coupling between the adiabatic basis states. This coupling is smallest in the initial and final times when the basis states are equally weighted superpositions of the diabatic basis states because $\theta(0) = \frac{\pi}{2}$. Meanwhile, the diabatic levels go through three crossings during the evolution. 

\begin{figure}[hb]
\begin{center}
\includegraphics[scale=0.6]{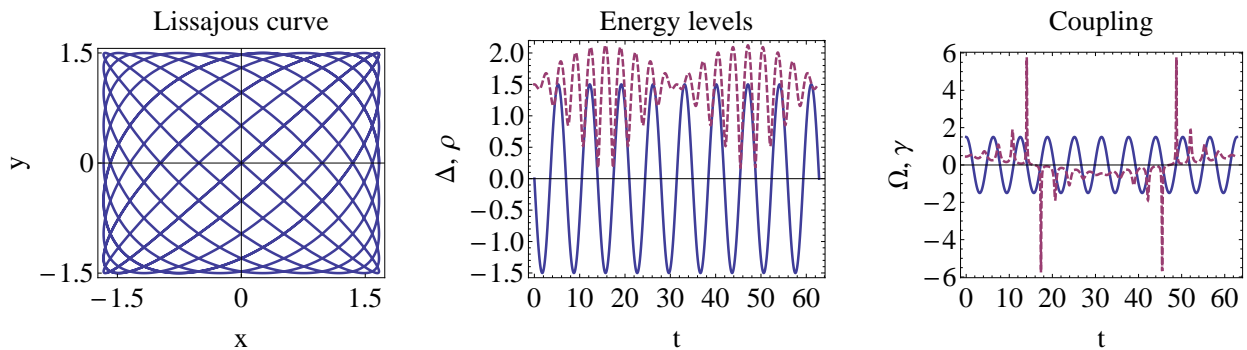} 
\caption{Left: An example of Lissajous curve of equation (\ref{eqn:lissajous}). Middle: Energy levels corresponding to Hamiltonian (\ref{eqn:Hamiltonian3}). Diabatic levels are depicted with a full blue line while the adiabatic levels are purple and dashed. Right: The diabatic (full line, blue) and the adiabatic (dashed, purple) coupling of (\ref{eqn:Hamiltonian2}). The parameters in all of the plots are $a = b = 1.5$, $\delta = \frac{\pi}{2}$ and $n = 0.9$. It should be noted that had the $n$ been equal to unity, these plots would be similar to the middle panel of figure \ref{fig:ellipsefunctions}.  }
\label{fig:lissajous}
\end{center}
\end{figure}

As a final example, we note that a Hamiltonian of the form
\begin{eqnarray}
\Delta_{3}(t) &= a\cos(nt + \delta )\\
\Omega_{3}(t) &= b\cos(t),
\label{eqn:Hamiltonian3}
\end{eqnarray}
where $a$, $b$, $n$ and $\delta$ are real parameters gives rise to the Lissajous curve
\begin{eqnarray}
x_{3}(t) &= \frac{a}{n}\sin(nt + \delta )\\
y_{3}(t) &= b\sin(t).
\label{eqn:lissajous}
\end{eqnarray}
Again, this curve has well-known properties and it can be in general quite complicated. The simple special cases include the circle $(n = 1, \: a = b, \: \delta = \frac{\pi}{2})$, a line $(n = 1, \: \delta = 0)$ and a parabola $(n = 2, \: \delta = \frac{\pi}{2})$. In any case it is bounded around the origin by a box with sides $\frac{2 a}{n}$ and $2 b$ and when $n$ is rational, $n=\frac{k}{l}$, it is closed and the corresponding Hamiltonian is periodic. The curve is nevertheless highly sensitive to the changes in the values of $n$ and the nominator $k$ and denominator $l$ give the number of "lobes" in vertical and horizontal directions, respectively. So arbitrarily small deviation of $n$ from unity will affect greatly to the Lissajous figure compared to the simple elliptical shape when $n = 1$, see figure \ref{fig:lissajous}. This means that small deviations in the angle frequencies of the different trigonometric functions in the Hamiltonian (\ref{eqn:Hamiltonian3}) give very different behaviors for the quantum system, given long enough time, as one can associate each  "lobe" with some maximum of the curvature.

\section{Discussion}
\label{sec:discussion}

We have discussed the general formulation of the dynamics of the time-dependent two-level quantum systems and pointed out how there is a natural correspondence in the theory of plane curves. It is remarkable that by simply defining components of a curve parametrically as integrals of the basic functions of Hamiltonian, namely detuning of the energy levels and coupling between the basis states, we get simple geometrical interpretations for the basic physical variables. Taking this point of view, adiabatic coupling is curvature, eigenenergy is the speed of the curve and the polar angle of the Hamiltonian in (\ref{eqn:hspherical}) is the turning angle of the curve, for example. Also the coherent adiabatic dynamics is completely given by the geometry of the curve in a clear and intuitive manner. Non-adiabatic transitions happen near the points where the curve bends most and crossings of the eigenenergies will induce singularities in the adiabatic coupling. To consider dynamics outside adiabatic region one can, for example, consider curves defined analogously in the higher-order superadiabatic bases of \cite{DreseHolthaus1998}, where the $n$th and $(n + 1)$th order bases take the role of the diabatic and adiabatic bases, respectively. 

One important consequence of this formulation is that now one can apply the mathematical results of the theory of plane curves to the dynamics of two-level systems in a direct way. The fundamental theorem of plane curves gives a constructive way to obtain interesting Hamiltonians with given adiabatic couplings. Also, we discussed how the four-vertex theorem gives conditions for the approach of the adiabatic limit in the case of Hamiltonians which correspond to simple closed curves. Unlike the curve of the parabolic model in figure \ref{fig:ex1} which in the adiabatic limit tends to a line, the total curvature of a closed plane curve is always a multiple of $2\pi$ so it can tend to the adiabatic limit somewhat differently, by smoothing the peaks in the curvature but also increasing in length. 

Many of the parametrizations of the closed plane curves involve trigonometric functions and we considered three different periodic level-crossing models motivated by the differential geometry of such plane curves. Each of these exhibit interesting features, for example, when it comes to parameter dependence and to the level structures and couplings when compared in different bases. These models were not directly obtained from a particular physical problem but this  is not a drawback since nowadays one can modify the time-dependencies of the external fields at will, with the advent of new laser technologies, for example. It should be also noted that many simple models can be obtained for certain special parameter values for the studied models. 

\ack % instead of \section{Acknowledgements}

This research was supported by the Finnish Academy of Science and Letters, and the Academy of Finland, grant 133682.

\appendix
\section*{Appendix}
\setcounter{section}{1}
We introduce here just the basic definitions and properties of plane curves that suffice to our purposes of relating them with time-dependent TLS as discussed in the main text. The definitions and notations here are mostly standard and more information can be found, for example, from \cite{GrayBook, Guggenheimer, Stoker}.

A parametrized curve is defined as a (piecewise-) differentiable function $\bi{\alpha}: I \rightarrow \mathbb{R}^{n}$, 
%\begin{equation}
%\bi{\alpha}: I \rightarrow \mathbb{R}^{n},
%\end{equation}
where $I = (a, b) $ is an open interval in $\mathbb{R}$, either finite or infinite. %The definition of the set $I \in \mathbb{R}$ can actually be a little bit more general but this suffices for our purposes. 
Furthermore, we often differentiate the curves freely so in effect we actually assume the curves to be smooth. As we are interested in curves in the plane, we set $n = 2$. The plane curve can be considered as a parametrized vector given in a Cartesian coordinates as 
\begin{equation}
\bi{\alpha}(t) = \left( x(t), y(t)\right),
\end{equation}
where $x(t)$ and $y(t)$ are two real functions and $t \in I$. The derivative of a curve is obtained naturally by componentwise differentiation and $\dot{\alpha}$ and $\ddot{\alpha}$ are called the velocity and acceleration of a curve $\alpha$, respectively. 
The norm of the velocity vector, $v(t) = \Vert \dot{\alpha} (t) \Vert$, is called speed. For a curve to be well-behaved it is required that $v (t) \neq 0$ for all $t$ and such curves are called regular.

Two distinct functions can trace the same point set on the plane, so it may not always be immediately clear when two curves are actually the same, i.e., they differ only by the parametrization. If we have two curves, $\alpha: (a, b) \rightarrow \mathbb{R}^{2}$ and $\beta: (c, d) \rightarrow \mathbb{R}^{2}$ and there exists a differentiable function $h: (c, d) \rightarrow (a, b)$ such that $\dot{h}(t) > 0$ ($\dot{h}(t) < 0$) for all $c < t < d$  and $\beta = \alpha \circ h$, we say that $\beta$ is a positive (negative) reparametrization of $\alpha$. The different signs of the reparametrization are related only to the direction the curve is traversed. From an intuitive geometrical point of view it is also clear that any purely geometrical quantity associated with curves should not depend nontrivially on the parametrization.

One such basic geometric quantity is the length $L[\alpha]$ of a curve $\alpha: (a, b) \rightarrow \mathbb{R}^{n}$ %(and it is assumed that $\alpha$ is defined in a slightly larger interval)
, defined by
\begin{equation}
L[\vec{\alpha}] = \int_{a}^{b}\Vert \vec{\alpha}'(u)\Vert du.
\end{equation}
%This obviously is independent of the parametrization, as follows straightforwardly from the chain rule and the change of integration variable. 
A closely related quantity is the arc-length function of curve. Fix a number $ c \in (a, b)$ and let the upper limit of integration be the variable $t$,
\begin{equation}
l[\vec{\alpha}, c](t) = \int_{c}^{\tau}\Vert \vec{\alpha} '(u)\Vert du.
\end{equation}

For any regular curve $\vec{\alpha} (t)$ there exists a reparametrized unit-speed curve $ \vec{\beta} (s) $, meaning that $\Vert \vec{\beta} ' (s) \Vert = 1$. The unit-speed parameter, which is essentially unique (up to change of origin and sign), is hereafter denoted by $s$. Because of the property $l[\vec{\beta}, c](s) = s - c$, unit-speed curves are said to be parametrized by the arc length. The unit-speed parametrization is very useful as many of the formulas simplify when it is used.

The most important quantity one can associate with a plane curve is its curvature $\kappa\left[\alpha\right](t)$. It basically measures the way the plane curve differs from a straight line, being identically zero only for a line and constant if and only if the curve is an arc of a circle. A formula for a regular curve is given by 
\begin{equation}
\kappa\left[\alpha\right](t) = \frac{\dot{x}(t)\ddot{y}(t) - \ddot{x}(t)\dot{y}(t)}{\left( \dot{x}^{2}(t) + \dot{y}^{2}(t)\right)^{3/2}}.
\label{eqn:curvature1}
\end{equation}
To see more clearly the meaning of the curvature function, one can associate two orthonormal vectors on each point of a unit-speed curve, the tangent vector $\bi{t}(s)$ and a vector obtained by rotating this by $\pi/2$, namely the normal vector $\bi{n}(s)$. Then the curve can be obtained from the Frenet equation
\begin{eqnarray}
\frac{\rmd}{\rmd s}\left( \begin{array}{c} \bi{t}(s) \\ \bi{n}(s) \end{array}\right) = \left(\begin{array}{lr} 0 & \kappa(s) \\
																						-\kappa(s) & 0	\end{array}\right) 
																						\left(\begin{array}{c} \bi{t}(s) \\\bi{n}(s)  \end{array}\right).
\label{eqn:freneteq}
\end{eqnarray}

The content of the fundamental theorem on plane curves is that curvature determines the plane curve essentially uniquely, meaning up to Euclidean motions and reparametrizations. A curve with any desired curvature can be realized with the unit-speed construction
\begin{eqnarray}
\alpha (s) &= \left( \int \cos\theta(s) ds + c, \, \int \sin\theta(s)ds + d \right) \nonumber \\
\theta (s) &= \int \kappa(s)ds + \theta_{0},
\end{eqnarray}
where $c$, $d$ and $\theta_{0}$ are constants. The function $\theta(s) $ is also called the turning angle of the curve and is the angle between the x-axis and the tangent vector of the curve at point $t$.

Finally, the four-vertex theorem states that the curvature function of any simple closed plane curve, other than a circle, must have at least four vertices, that is, points where the curvature is locally minimal or maximal. One should also note that if we relax the condition of simplicity, the curve can have less than four vertices.


\section*{References}
\begin{thebibliography}{50}

\bibitem{Berry263} Berry M V 1995 \textit{Ann. N.Y. Acad. Sci.} \textbf{755} 303

\bibitem{Shore2011} Shore B W 2011 \textit{Manipulating Quantum Structures Using Laser Pulses} (Cambridge: Cambridge University Press)

\bibitem{Stenholm2005} Stenholm S and Suominen K-A 2005 \textit{Quantum Approach to Informatics} (Hoboken: John Wiley \& Sons)

\bibitem{chruscinski} Chru\'{s}ci\'{n}ski D and Jamiolkowski A 2004 \textit{Geometric Phases in Classical and Quantum Mechanics} (Boston: Birk\"{a}user)

\bibitem{Berry201} Berry M V 1990 {\it Proc. R. Soc. A} \textbf{429} 61

\bibitem{Zener1932} Zener C 1932 {\it Proc. R. Soc. Lond. A} \textbf{137} 696

\bibitem{Landau1932} Landau L D 1932 {\it Phys. Z. Sowjet Union} \textbf{2} 46 

\bibitem{Stuckelberg1932} St{\"u}ckelberg E C G 1932 {\it Helv. Phys. Acta} {\bf 5} 369

\bibitem{Majorana1932} Majorana E 1932 {\it Nuovo Cimento} \textbf{9} 43; Bassani G F (ed.) 2006 \textit{Ettore Majorana: Scientific Papers} (Bologna: SIF)

\bibitem{Rojo2010} Rojo A G and Bloch A M 2010 {\it Am. J. Phys.} \textbf{78} 1014

\bibitem{Berry242} Berry M V and Robbins J M 1993 {\it Proc. R. Soc. A} \textbf{442} 641

\bibitem{Berry206} Berry M V 1990 {\it Proc. R. Soc. London A} \textbf{430} 405%geometric amplitude factors in quantum transitions

\bibitem{Vitanov2001} Vitanov N V, Halfmann T, Shore B W and Bergmann K 2001 \textit{Annu. Rev. Phys. Chem.} {\bf 52} 763

\bibitem{SuominenParabolic} Suominen K-A 1992 {\it Opt. Comm.} \textbf{93} 126

\bibitem{LehtoSuominen2012} Lehto J and Suominen K-A 2012 {\it Phys. Rev. A} \textbf{86} 033415

\bibitem{Lehto2013} Lehto J 2013 {\it Phys. Rev. A} \textbf{88} 043404

\bibitem{GrayBook} Gray A, Abbena E and Salamon S 1997 \textit{Modern Differential Geometry of Curves and Surfaces with Mathematica, 3rd ed.}, (Boca Raton: Chapman and Hall/CRC)

\bibitem{DreseHolthaus1998} Drese K and Holthaus M 1998 \textit{Eur. Phys. J. D} {\bf 3} 73

\bibitem{Guggenheimer} Guggenheimer H W 1963 \textit{Differential Geometry}, (New York: McGraw-Hill Book Company)

\bibitem{Stoker} Stoker J J 1989 \textit{Differential Geometry}, (New York: John Wiley \& Sons)

\end{thebibliography}
\end{document}